# Brief Studying of Oil Crust Thickness Measurement by Gamma Ray Compton Scattering Approach

Mamat.Mamatrishat[1] , Aierken.Abuliemu[1], Ding Jie [2] Wang Shiheng[1],

(1. Physics department of Xinjiang University, Urumqi, Xinjiang, China 830046)
(2. Physics department of Nanjing University, Nanjing, Jiangsu, China 210093)

**Abstract:** The relation between the scattering cross section and the scattering angle under different energy condition of the incident rays is analyzed. From Compton scattering total cross section, a formula of quasi-parallel incident gamma ray Compton scattering response function versus to thickness of oil crust target is derived and analyzed. Numerical fitting result shows that there exists cubic relation between response function of gamma ray and thickness of oil crust.

Ⅰ. **Introduction**

It's very important to measure oil crust thickness of oil pipeline without destroying pipelines and to keep certain amount of oil flow in the tube while oil is transporting by the tube. Gamma Ray Compton Scattering Method (GRCSM) is a well known approach to detect and measure surface defects in materials [1-5]. Using Gamma Ray (GR) to determine the oil crust is having great applications not only because of its practical availability, but also its low economic cost priority. There are many theoretical and experimental works done by the Gamma Ray Transmission Method (GRTM) which measure the material characters [6-8], and both the theoretical and experimental results exhibit good agreement with each other. In our previous work [9], we have measured oil crust thickness by GRTM, but for the underground imbedded oil pipelines, it is interesting to use GRCSM. In this work, first we analyze the relation between the incident rays and the scattered rays according to Compton Scattering (CS) theory. Meanwhile we discuss the intensity of gamma ray versus to the scattering angle. Three different kinds of approach (GRTM, large angle which is backward scattering approach and small angle scattering approach by quasi-parallel incident) of CS are separately discussed. Finally we give a cubic relation between the response function of GR and thickness of oil crust that exist in the quasi-parallel incident small angle CS approach. This might be very useful in choosing proper experimental parameters in real measurements.

Ⅱ. **Compton scattering and scattering cross section**

In order to compare the future experimental result with the theoretical result, we consider CS of GR with the energy of 0.6616 MeV from $^{137}$Cs into oil crust target. Generally, bounded electrons of the target nucleus can be considered as the free electrons because the bounding energy of the outshell electrons is far less than the energy of 0.6616 MeV. When CS happens, the momentum and energy transform

**Received date:**
**Foundation item:** Supported by National Natural Science Foundation of China (No.10565003), and supported by UIJ of Xinjiang University (070195).
**Biography:** Mamat Mamatrishat (1975-), male, lecturer.
**Corresponding author:** Aierken.Abuliemu (1956-), male, Professor.

between the GR photons and electrons of the target medium, as the result, the energy and the wavelength of the scattering photon are changed. Assuming $E_0 = h\nu_0$, $E_\theta = h\nu_\theta$ are the initial energy of the incident photon and the final energy of the scattered photon after colliding with the electron respectively, and $\theta$ is an angle between the incident photon and the scattering photon rays, $\varphi$ is an angle between the scattering photon and the recoiling electron rays (Fig.1 illustrates CS process). From the relativistic energy and the momentum conservation relations, we obtain

$$E' = \frac{E_0}{1 + \frac{E_0}{m_e C^2}(1-\cos\theta)} \quad (1)$$

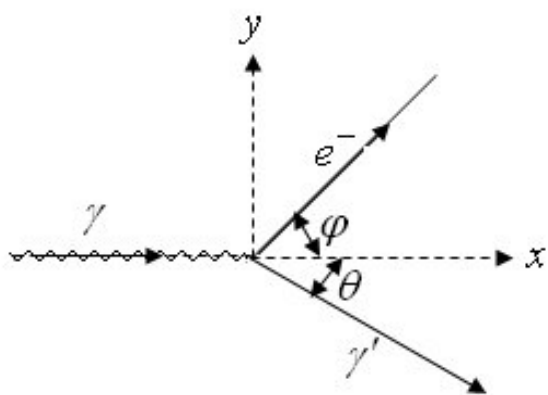

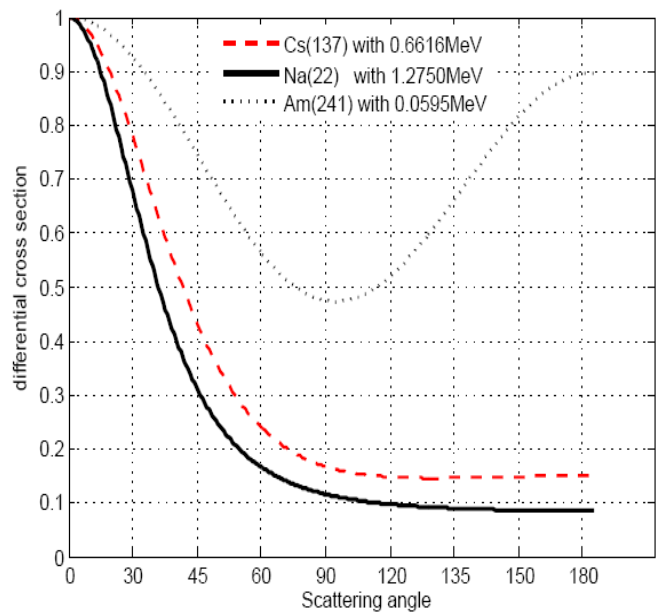

**Fig. 1** Compton scattering process

**Fig.2** Differential cross section of Compton scattering as a function of scattering angle for incident photons with different energies.

CS of the photon is a process of the photon and the electron interaction. In 1928, Klein and Nishina showed the quantum mechanical CS from relativistic Dirac theory of the electron. From Klein-Nishina equation [10], the scattering cross section of the CS is written as

$$\frac{d\sigma}{d\Omega} = r_0^2 \frac{1+\cos^2\theta}{2} \cdot \frac{1}{[1+\gamma(1-\cos\theta)]^2} \cdot \{1 + \frac{\gamma^2(1-\cos\theta)^2}{(1+\cos^2\theta)[1+\gamma(1-\cos\theta)]}\} \quad (2)$$

where $r_0 = e^2/m_e c^2 = 2.82 \times 10^{-13}$ is the classic radius of the electron, and $\gamma = h\nu_0/m_e c^2$ is the relativistic factor of the photon, for radioactive source $^{137}$Cs is $m_e c^2$=1.2947.

Fig. 2 shows the relation of $d\sigma/d\Omega$ as the function of $\theta$, when $\theta$ changes from $\theta$ to $\theta + d\theta$. We see that $d\sigma/d\Omega$ is almost constant when the scattering angle is larger than $\frac{\pi}{2}$ for $\gamma$ ray photon with the energy of 0.6616 MeV. The value of $d\sigma/d\Omega$ is about 10% of maximum value of $d\sigma/d\Omega$. Therefore $\gamma$ ray photons with the energy of 0.6616 MeV, large scattering angle ($\theta > \frac{\pi}{2}$) might be a good solution to measure the scattering photon.

### III. Oil crust measurement by $\gamma$-ray transmission method

In $\gamma$-ray transmission method, the incident GR mostly irradiate to the target surface (Fig. 3-a) rectilinearly. The intensity of GR after it transmitted to the target with the thickness of $X$ is

$$I(X) = I_0 e^{-n_e \sigma_t X} \quad (3)$$

where, $I_0$ is the initial intensity of the incident GR, $I(X)$ is the intensity of the transmitted GR, $n_e$ is the electron density in the target medium, $\sigma_t$ is the total cross section.

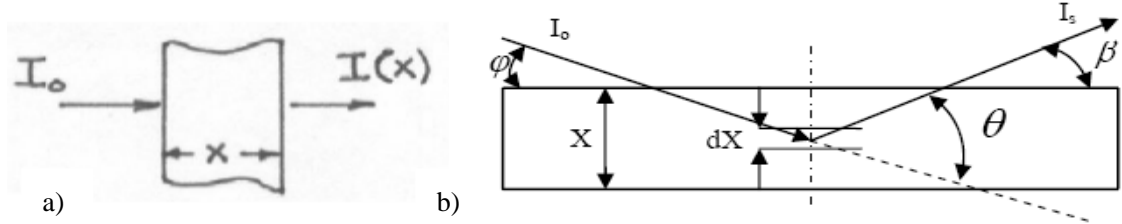

**Fig. 3** Gamma ray scattering: a) transmission method, b) small angle scattering method by the quasi-parallel incident

Using equation (3), we obtain

$$X = \frac{1}{n_e \sigma_t} \ln \frac{I_0}{I} \quad (4)$$

The response function $\ln \frac{I_0}{I}$ in Eq. (4) can be measured by the experiment, $\sigma_t$ is the total cross section, and $n_e$ is known from the chemical gradients of the target medium (i.e., the oil crust). Therefore in principle the thickness of the oil crust can be measured by the transmission method once one knows $\sigma_t$, $n_e$ and $\ln \frac{I_0}{I}$. Because the decay coefficient of the medium is defined as $\mu = \sigma_t n_e$, the thickness of the medium is also analyzed by using $\mu$. Fig. 4 shows a linear relation between the response function $\ln\frac{I_0}{I}$ and the thickness of the target X. Here, we have taken the decay coefficient as 0.00548 mm$^{-1}$, which is measured in our previous work [9] of measuring oil crust thickness by GRTM.

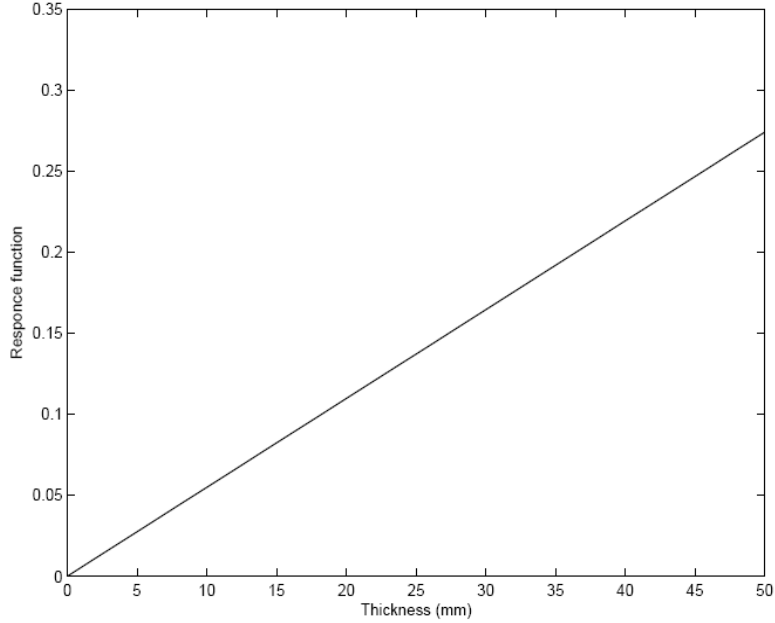

**Fig.4** Response functions of the Compton scattering as a function of the target thickness, the decay coefficient is 0.00548 mm$^{-1}$.

### IV. Measuring the oil crust thickness by the rectilinear irradiant large scattering angle

In order to find a proper angle of inverse CS, in Fig. 2 we illustrate the relation between the scattering angle and the scattering cross section for the photons with the different energies from the Klein-Nishina equation. For the GR with the energy of 0.6616 MeV (or the energy with 1.2750 MeV from $^{22}$Na source), we see that the differential cross section is decreased while the incident ray photon energy is increased, i.e., for the larger energy of the incident ray photon, there is the smaller value of CS cross section. Meanwhile the differential cross section is decreased while the scattering angle is increased, i.e., the scattering cross section which corresponds to the large scattering angle is much smaller than the acute angle scattering cross section.

Fig. 2, shows that the value of the differential cross section is strongly related to the energy of the incident ray photon in the direction of the backward scattering (i.e., $\theta \geq \frac{\pi}{2}$). For the incident photons with the energy of 0.6616 MeV, the scattering cross section is tends to vanish when the scattering angle is larger than the angle $\frac{\pi}{2}$. Similarly the scattering cross section is kept a constant while the scattering angle is increased. Therefore, for the incident photons with the energy of 0.6616 MeV, we don't select a proper angle in the inverse CS (or large angle CS) case.

If we decrease the energy of the incident photon (choosing $^{241}$Am as an irradiative source with the photon energy of 0.0595 MeV) or quasi-parallel irradiate GR on the target surface, we can find the proper scattering angle to measure the counting numbers of the photons. For example, if we use GR with the energy of 0.0595 MeV from source $^{241}$Am, we find that the scattering cross section is increased linearly while scattering angle is increased in the region of the large angle ($\theta > \frac{\pi}{2}$, see at Fig. 2), but the value of the scattering cross section corresponding to the inverse scattering is still smaller than the value of the scattering cross section of the scattering angle $0 \sim 30^o$.

When the energy of the incident photon goes to zero, the value of the scattering cross section corresponding to the large scattering angle is increased efficiently. However at the same time, because the energy of the incident photon is decreased, we may not be able to more accurately measure the thickness of the target medium.

Ⅴ. **The small angle scattering by the quasi parallel incident**

If the incident GR is quasi-parallel to the surface of the target medium (Fig. 3-b), then we are able to measure the scattering GR with the large cross section value in the regime of the acute angle scattering. Assuming a bound of GR incident to the thickness $X'$ of the target medium by the angle $\varphi$, that is an angle between the incident $\gamma$-rays and the surface of the target medium, we label the intensity of the incident GR as $I_0$, and the intensity of the scattered GR measured by detector as $I_S$. And $\beta$ is an angle between the scattering GR and the surface of the target medium. If the target medium is isotopic, then the decay coefficients $\mu = n_e \sigma_t$ of different directions are homogenous.

The intensity of the transmitted GR in the depth x, is obtained from Eq. (3)

$$I_t(x) = I_0 e^{-\mu(\frac{x}{\sin\varphi})} . \quad (5)$$

Then the GR transmitted over the depth x is scattered by the target atoms in the thickness of $dx$, and the scattered GR intensity is

$$dI_S(x) = I_t(x)\sigma_t n_e \frac{dx}{\sin\varphi}\Omega(x) , \quad (6)$$

where $\Omega(x)$ is a solid angle. After scattering, the intensity of the GR transmitted to the direction of angle $\beta$ is

$$I_s(x) = \int_0^{X'} I_0 n_e \sigma_t \Omega(x) \exp[-\mu(\frac{1}{\sin\varphi} + \frac{1}{\sin\beta})x]\frac{1}{\sin\varphi}dx . \quad (7)$$

When the distance between the detector and the target surface is large enough comparing to the thickness of the target medium, $\Omega(x)$ can be considered as a constant. Then Eq. (7) can be integrated as

$$\frac{I_s}{I_\circ} = \Omega\{1 - \exp[-\mu(\frac{1}{\sin\varphi} + \frac{1}{\sin\beta})X']\}/(1 + \frac{\sin\varphi}{\sin\beta}) . \quad (8)$$

Fig. 5 shows the relation between the intensity of the scattering GR and the thickness of the target medium. Here, we take the decay coefficient as same as in the case of GRTM, $\varphi = \beta = 30°$, and $\Omega = \pi/18$. After polynomial fitting the curve which is numerically obtained from Eq. (8), we found that the cubic fitting is the best one. Therefore the cubic increasing relation of the intensity versus to the thickness is satisfied in the case of small angle scattering by the quasi-parallel incident.

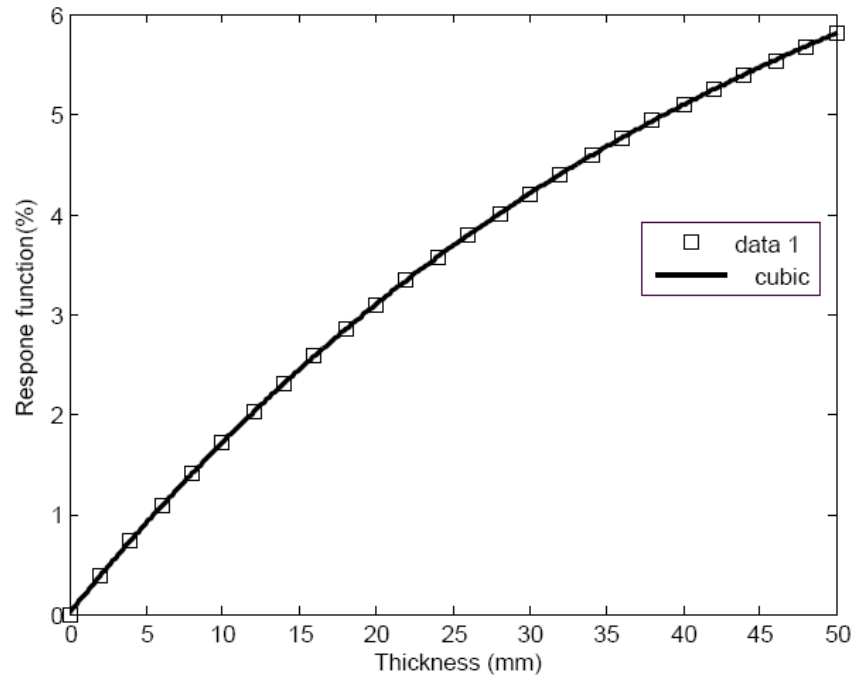

**Fig.5** Response functions of Compton scattering as a function of the target thickness, the □ line is numerical curve of the Eq. (8), and the solid line curve is a cubic fitting result.

## Ⅴ.Conclusion

To measure the thickness of the oil crust in the underground imbedded oil pipelines, we use small angle scattering by quasi-parallel incident approach in order to get higher intensity of the scattering ray. Small angle scattering by the quasi-parallel incident approach has great application to comparing with rectilinear irradiant approach once one of the sides of the target is inaccessible and the transmission approach is not applicable. This approach may be applied to the other kinds of GR measurements by CS.

**Acknowledgement**

We would like to thank China National Science Research Fond (Grant number: 10565003), Xinjiang University Research Fond (Grant number: 070195) for the financial support, and we are grateful to Professor Liu Shengkang for his kind discussion and guidelines. We thank Dr. Wuernisha for her corrections of English.

**Corresponding author:** Aierken.Abuliemu, E-mail: ark@xju.edu.cn，Phone:+86-991-8212304 Fax:+86-991-8582405